# On the Importance of Performing App Analysis Within Peer Groups

Safwat Hassan
*Dept. of Engineering*
*Thompson Rivers University*
Kamloops, BC, Canada
shassan@tru.ca

Heng Li
*Dept. of Computer and Software Engineering*
*Polytechnique Montréal*
Montréal, QC, Canada
heng.li@polymtl.ca

Ahmed E. Hassan
*Software Analysis and Intelligence Lab (SAIL)*
*Queen's University*
Kingston, ON, Canada
ahmed@cs.queensu.ca

*Abstract*—The competing nature of the app market motivates us to shift our focus on apps that provide similar functionalities and directly compete with each other (i.e., *peer apps*). In this work, we study the ratings and the review text of 100 Android apps across 10 peer app groups. We highlight the importance of performing peer-app analysis by showing that it can provide a unique perspective over performing a global analysis of apps (i.e., mixing apps from multiple categories). First, we observe that comparing user ratings within peer groups can provide very different results from comparing user ratings from a global perspective. Then, we show that peer-app analysis provides a different perspective to spot the dominant topics in the user reviews, and to understand the impact of the topics on user ratings. Our findings suggest that future efforts may pay more attention to performing and supporting app analysis from a peer group context. For example, app store owners may consider an additional rating mechanism that normalizes app ratings within peer groups, and future research may help developers understand the characteristics of specific peer groups and prioritize their efforts.

## I. INTRODUCTION

Prior studies on app analysis usually focus their analysis on popular apps, aiming to cover apps from different domains and categories [1–5]. Such *global analysis* (i.e., analyzing apps from multiple categories) provides a holistic view of the commonly repeated issues across the studied apps. However, comparing apps that provide different functionalities may not spot the unique characteristics (e.g., the critical issues) of the closely related apps (i.e., *peer apps*). For example, even though the *"Firefox Browser"* app and the *"Skype"* app are both in the *"Communication"* app category in the Google Play Store, analyzing these two apps together may lead to noise in understanding the main challenges and the most important aspects of developing browser or telecommunication apps, as their major functionalities are very different. In reality, app users only compare closely related apps: apps that provide similar functionalities. Hence, app stores, such as the Google Play Store, have recently introduced features to enable developers to compare their app with a custom-defined peer group that contains closely related apps [6].

In this paper, we name apps that provide similar functionalities and directly compete with each other as **peer apps**, and we name a group of peer apps as a **peer group**. For example, *"The Weather Channel"* and *"AccuWeather"* are peer apps as they both provide similar functionalities (e.g., weather forecasting). On the other hand, the *"Firefox Browser"* app and the *"Skype"* app are not peer apps even though they are both in the same app category (i.e., *"Communication"*).

Prior studies propose different approaches to cluster similar apps in app stores [7–10]. In this work, we perform an in-depth study to understand the importance of performing app analysis within peer groups. Our *goal* is to demonstrate that analyzing peer apps provides a unique perspective from analysing a collection of unrelated apps as is commonly done today in literature [4, 11–18]. To eliminate any bias in our results with respect to a particular app group, we study peer apps across ten groups. We analyze the ten most popular apps for each peer group. In total, we analyze 100 apps across ten peer groups. These apps received a total of 6,773,653 reviews during our study period. Our work demonstrates the importance of performing peer-app analysis instead of performing a global analysis of apps (i.e., considering unrelated apps) along three research questions (RQs):

**RQ1:** *How does comparing user ratings within peer apps differ from comparing user ratings globally across all the apps?*

We observe that comparing user ratings within peer apps can provide very different results from comparing user ratings from a global perspective. For example, the lowest-rated app in one peer group (e.g., *"Bible"* apps) might have a higher rating than the highest-rated app in another peer group (e.g., *"Weather"* apps).

**RQ2:** *How does the peer analysis of user reviews differ from the global analysis?*

We find that review topics are mentioned heterogeneously in the reviews of the apps across different peer groups while being mentioned homogeneously in the reviews of the apps within the same peer groups. Peer-group analysis provides a complementary perspective to spot the dominant topics in the user reviews of a peer group. For example, UI-related and performance-related topics are the most dominant topics in the reviews of the *"Wallpaper"* and the *"Browser"* peer apps, respectively. However, such topics are not the most frequent ones across all the studied apps.

**RQ3:** *How do review topics contribute to the negative ratings within peer apps versus globally?*
We find that the same topics have a substantially different contribution to the ratings of an app across peer groups, while the same topics have a similar contribution to the ratings of the apps within the same peer groups. A seemly "less harmful" topic from a global perspective might be much more harmful to certain peer apps.

Our findings provide the following implications:

1) **App store owners** may consider providing an additional rating mechanism that normalizes app ratings within peer groups, to provide app developers and users a different perspective about the position of an app.
2) **App developers**, in particular, of apps with fewer reviews, may prioritize their efforts to improve their apps or solve issues based on the most important aspects (i.e., users' main concerns) of their peer group.
3) **Software engineering researchers and tool developers** may consider peer-app analysis to help app developers understand the characteristics of specific peer groups and prioritize their efforts.

**Paper organization.** The rest of this paper is organized as follows. Section II describes our process of preparing data for our analysis. Section III presents our results for answering the proposed research questions. Section IV discusses the threats to the validity of our findings. Section V summarizes prior work that is related to our work. Finally, Section VI concludes our paper.

## II. EXPERIMENT SETUP

In this section, we describe our process of preparing data for our analysis. Figure 1 shows our process of selecting apps and extracting app data.

### A. Identifying Peer Apps

To study the impact of analyzing apps from the perspective of peer groups, we selected ten peer app groups which represent a broad range of app domains as follows:

**Step 1: Selecting popular Android apps.** We first selected the top 2,000 popular apps in the Google Play Store according to the App Annie report [19].

**Step 2: Identifying peer groups and peer apps.** We identify peer groups and peer apps within each peer group through examining the titles and descriptions of the 2,000 popular apps. Two authors of the paper manually read the title and description of each app to identify peer apps. When there is a conflict, the two authors discuss and reach an agreed-upon result. Apps are assigned to the same peer group when they provide similar major functionalities. For example, the *"The Weather Channel"* app and the *"AccuWeather"* app are assigned in the same group because they provide similar major functionalities: weather report and forecasting.

**Step 3: Filtering peer groups.** After identifying peer groups and apps within every peer group, we randomly selected ten peer groups that have at least ten apps. We studied different peer groups (i.e., ten groups) to ensure that our results are not biased towards a specific peer group. In addition, some peer groups have less than ten peer apps (i.e., a smaller number of popular competitors on the market). We did not consider such small peer groups as the characteristics of such small peer groups might be biased towards a small number of apps.

For every peer group with ten or more peer apps, we choose the ten most popular apps that have the largest number of reviews. We focus our study on the top popular apps as these apps contain rich review data that facilitate our analysis of app ratings (RQ1) and reviews (RQ2 and RQ3). In total, we study 100 apps that fall into 10 peer groups. Table I lists all the studied apps and their peer groups.

### B. Collecting App Data

We used a web crawler [20] to collect the data of our studied apps for 21 months. We extracted the general information about each app, including the app description and the average rating during our crawling period. The data of the studied apps were crawled on a daily basis to ensure that the complete history of the studied apps in the app store was captured. We leverage the dynamic information of each app, for example, to analyze the changes of app ratings over time (in RQ1).

We also extracted each user review of the studied apps, including the review time, the review text and the corresponding rating. Table III shows a summary of the number of reviews that we collected for each studied app. Table II shows the number of reviews that we collected for each peer group. In total, we collected 6,773,653 reviews for the studied apps.

### C. Extracting Sample Reviews

As shown in Table III, different apps may have a substantially different number of user reviews (e.g., up to three magnitudes in difference, as shown in the row "# Collected Reviews"). Such substantial difference also propagates to peer groups. As shown in Table II, the number of collected reviews for each peer group has up to eight times in difference. Therefore, our analysis of review topics (in RQ2 and RQ3) tends to be biased towards the apps and the peer groups with a larger number of reviews. The apps or peer groups with a larger number of reviews are likely to dominate the results of the extracted review topics [21].

To avoid such a bias, when we extract review topics, we randomly sample a statistically representative sample of reviews from each studied app. Table III summarizes the number of randomly sampled reviews per app. Our sample for each studied app range from 444 to 2,392 which represents the overall reviews of each app with a confidence level of 98% and a confidence interval of 5%. The difference between the number of studied reviews for each app is much lower after our sampling process. In total, our random sample contains 204,835 reviews across 100 studied apps. Table II shows the number of randomly sampled reviews for each peer group, which ranges from 17,529 to 23,412. The balanced distribution of user reviews across peer groups ensures that the results of our analysis are not biased towards certain peer groups.

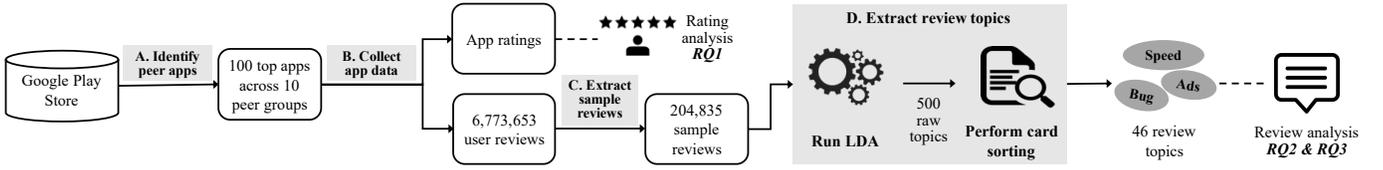

Fig. 1. Our process for selecting apps and extracting app data.

TABLE I
THE STUDIED PEER GROUPS AND APPS IN EACH PEER GROUP.

| Peer Group | Primary Functionality | Apps |
|---|---|---|
| Weather | Weather report and forecast | 1. Weather & Clock Widget; 2. AccuWeather; 3. The Weather Channel; 4. GO Weather; 5. Yahoo Weather; 6. Weather; 7. Weather by WeatherBug; 8. Transparent clock & weather; 9. Weather Live Free; 10. 1Weather |
| Bible | Bible content | 1. Bible; 2. Bible Offline; 3. King James Bible (KJV) Free; 4. Daily Bible; 5. Bible App for Kids; 6. Bible: Dramatized Audio Bibles; 7. Superbook Bible, Video & Games; 8. Holy Bible King James + Audio; 9. Audio Bible MP3; 10. Bible KJV |
| Browser | Web browsing | 1. UC Browser; 2. Opera Mini 3. Firefox Browser; 4. Dolphin Browser; 5. Opera browser; 6. CM Browser; 7. Puffin Web Browser; 8. Web Browser & Explorer; 9. Photon Flash Player & Browser; 10. Adblock Browser for Android |
| Navigation | Maps and GPS Navigation | 1. Google Maps; 2. Waze, 3. GPS Navigation & Offline Maps Sygic, 4. MapFactor GPS Navigation Maps; 5. MAPS.ME; 6. Free GPS Navigation; 7. HERE WeGo; 8. Maps, GPS Navigation & Directions, Street View; 9. Offline Maps & Navigation; 10. Scout GPS Navigation & Meet Up |
| Free call | Free calls and instant messaging | 1. Skype; 2. LINE; 3. imo; 4. KakaoTalk; 5. BBM; 6. free video calls and chat; 7. Free phone calls, free texting; 8. TalkU Free Calls +Free Texting +International Call; 9. Talkatone: Free Texts, Calls & Phone Number; 10. WePhone - free phone calls & cheap calls |
| SMS | Supporting SMS service | 1. Truecaller; 2. GO SMS Pro; 3. Textra SMS; 4. SMS from PC / Tablet & MMS Text Messaging Sync; 5. Truemessenger - SMS Block Spam; 6. Nextplus Free SMS Text + Calls;7. Block call and block SMS; 8. Messenger - SMS, MMS App; 9. Handcent Next SMS; 10. Messages + SMS |
| Music Player | Playing music | 1. Poweramp Music Player; 2. Music Player (by Leopard V7); 3. Music Player (by mytechnosound); 4. SoundHound; 5. PlayerPro Music Player; 6. Free Music MP3 Player; 7. Equalizer music player booster; 8. Vevo - Music Video Player; 9. Music Player (by JRT Studio Music Apps); 10. Music - Mp3 Player |
| News | Providing news content | 1. Flipboard; 2. Google News; 3. Reddit; 4. News Republic; 5. BBC News; 6. CNN Breaking News; 7. Google News & Weather; 8. Fox News; 9. SmartNews; 10. Yahoo News |
| Security | Antivirus and Space Cleaner | 1. Clean Master; 2. Security Master; 3. 360 Security; 4. AVG AntiVirus; 5. DFNDR Security; 6. Avast Antivirus; 7. Power Clean; 8. 360 Security Lite; 9. GO Security; 10. Norton Security and Antivirus |
| Wallpaper | Themes and Wallpapers | 1. GO Launcher ; 2. ZEDGE Ringtones & Wallpapers; 3. CM Launcher 3D ; 4. APUS Launcher ; 5. Hola Launcher; 6. WhatsApp Wallpaper; 7. Backgrounds HD (Wallpapers); 8. GO Locker; 9. ZenUI Launcher; 10. Icon wallpaper dressup |

TABLE II
THE NUMBER OF COLLECTED REVIEWS AND SAMPLED REVIEWS FOR EACH PEER GROUP.

| Peer Group | Security | Browser | Wallpaper | Free call | SMS | Navigation | Weather | News | Music player | Bible | Total |
|---|---|---|---|---|---|---|---|---|---|---|---|
| # Collected Reviews | 1,422,233 | 1,071,999 | 906,160 | 860,347 | 784,716 | 640,472 | 395,539 | 290,611 | 213,954 | 187,622 | 6,773,653 |
| # Sampled Reviews | 23,412 | 19,561 | 22,276 | 21,315 | 18,285 | 19,522 | 21,746 | 21,223 | 19,966 | 17,529 | 204,835 |

TABLE III
MEAN AND FIVE-NUMBER SUMMARY OF THE NUMBER OF COLLECTED REVIEWS AND THE RANDOMLY SAMPLED REVIEWS PER APP.

|  | Min | 1st Qu. | Median | Mean | 3rd Qu. | Max |
|---|---|---|---|---|---|---|
| # Collected Reviews | 544 | 8,164 | 27,640 | 67,736 | 73,434 | 612,723 |
| # Sampled Reviews | 444 | 1,856 | 2,209 | 2,048 | 2,325 | 2,392 |

### D. Extracting Review Topics (LDA & Card Sorting)

As shown in Table II, the studied apps have a total number of 6,773,653 reviews that are posted during our studied period. After randomly sampling the reviews (Section II-C), there are still 204,835 reviews. It is almost impossible to manually go over all these reviews to understand how people perceive the studied apps. Therefore, we use automated topic modeling to extract the high-level topics of user reviews.

Determining the appropriate number of topics is a known challenge for studies that leverages automated topic modeling (e.g., LDA [22]). Prior work determines the number of topics either based on researchers' experience and manual experiments [23–25] or through the use of automated approaches to find the optimal number of topics [26]. However, existing approaches for searching the optimal number of topics are heuristic-based. As shown in prior work [26], different approaches (e.g., [27, 28]) can produce very different optimal numbers of topics. In this paper, we combine automated topic modeling and manual analysis to extract meaningful topics from the review text. Our approach leverages both the power of automated topic extraction and human insights. Figure 1 illustrates our topic extraction process. We first run LDA using a relatively large number of topics (i.e., 500 topics), as suggested by prior work [24, 29, 30]. Then, we perform a card sorting process to manually group similar topics together. Such a combination helps us identify more meaningful topics than only running LDA with a smaller number of topics. Prior works have used a similar combination to manually merge automatically generated topics that are semantically similar [25, 26]. In the next sections, we describe the detailed steps for our semi-automated approach of running LDA then performing card sorting.

*1) Running LDA:* We treat each sampled user review as a document and apply topic modeling on the user reviews to derive review topics. We use Latent Dirichlet allocation (LDA) [22] to derive topics from user reviews[1]. In LDA, a *topic* is a collection of frequently co-occurring words in the corpus. Given a corpus of $n$ documents $f_1, ..., f_n$, LDA automatically discovers a set of topics $Z$, where $Z = \{z_1, ..., z_K\}$, as well as the mapping $\theta$ between the topics and the documents. We use the notation $\theta_{ij}$ to describe the *topic membership* value of topic $z_i$ in document $f_j$. Formally, each topic is defined by a probability distribution over all of the unique words in the corpus. The number of topics, $K$, is an input that controls the granularity of the topics. In this work, we choose $K = 500$ as our number of topics. Prior work [24, 29, 30] suggests that using a larger number of topics has a lower risk than using a smaller number of topics, as the additional topics would have low topic membership values (i.e., noise topics) and can be filtered out (something that our follow up manual card sorting process can perform easily).

*2) Performing Card Sorting:* Automated topic modeling (e.g., LDA) tends to generate indistinct topics, even with a small number of topics [25, 32]. Therefore, we manually examined the resulting 500 topics and used an open card-sorting method [33, 34] to group similar topics together. We use our human insights and experience to identify similar topics if the most probable words of two topics are similar.

Two authors of the work (i.e., *coders*) jointly performed the open card-sorting process. We first printed each topic (i.e., the top 20 words associated with a topic) on a piece of paper (a.k.a., a card), then performed our open card sorting process in two broad phases:

**Phase-I sorting:** In this phase, following prior work [25, 26], we grouped similar topics together (i.e., topics with similar words or similar semantic meanings). For example, if two topics are both about "button", we grouped them together. The two coders jointly examined every topic. For each examined topic, we compared it with the previously derived groups. If we could not find an appropriate group for the examined topic, we created a new group and assigned the examined topic to the new group. If we don't understand a topic from the top 20 words, we check the reviews with the highest membership of the topic. We kept communicating during the entire sorting process, and we made decisions together. We constantly made changes to our existing groups whenever appropriate. The two coders spent around 30 hours together in this phase (across five sessions). As a result, we derived 138 topic groups after this phase.

**Phase-II sorting:** In this phase, we merged the lower-level topic groups, derived from phase I, into higher-level groups (i.e., themes). For example, we merged the topics about

[1]We use the MALLET implementation [31] of LDA.

"button" and the topics about "menu" into a higher-level topic group "UI design". The two coders jointly examined every lower-level topic group, following the same process as stated in "Phase-I sorting". The two coders spent around twelve hours in this phase (across two sessions). As a result, we derived 46 higher-level topic groups from the 138 groups derived in Phase I. Since the two coders examined all topics together, and agreements were reached for each topic, we did not compute the inter-rater agreement. In the rest of the paper, we use the term *"topic"* to refer to our manually-derived higher-level topic groups (i.e., the 46 groups derived in the Phase-II sorting).

### III. EXPERIMENT RESULTS

In this section, we demonstrate the motivation, approach, and results of our three research questions.

*A. RQ1: How does comparing user ratings within peer apps differ from comparing user ratings globally across all the apps?*

*1) Motivation:* The rating of an app plays a critical role for users who are deciding whether to download the app. Prior studies analyze app ratings from a global perspective (i.e., by randomly selecting subject apps that provide heterogeneous functionalities). However, users select apps relative to other apps that provide similar functionalities (e.g., weather forecasting) [8]. Therefore, in this RQ, we examine the difference between comparing apps within peer groups versus comparing apps irrelevant of their peer groups.

*2) Approach:* We wish to understand whether the analysis of app ratings within peer groups can provide new perspectives than a global analysis of app ratings. In this RQ, we analyze the user ratings of the studied apps along two measures: (1) average rating and (2) the variation of app ranks.

**Average ratings.** The Google Play Store provides the average rating of each app over all the reviews for the app across its whole history. For each studied app, we collected the information about its average rating from the Google Play Store at the end of the studied period.

We use the Kruskal-Wallis test [35] to evaluate the group differences of the average ratings: the average ratings of the studied apps is the response variable and their peer groups is the explanatory variable. The Kruskal-Wallis test is a non-parametric alternative of the one-way *analysis of variance test (ANOVA)*. We use the Kruskal-Wallis test as the average ratings of the studied apps are not normally distributed (i.e., the Shapiro-Wilk test [36] on the average ratings shows a p-value less than 0.05).

**Variation of app ranks.** While the average rating of an app indicates users' overall satisfaction with an app, it is the rank of an app (in particular, relative to its peers) that usually impacts users' choices. In this RQ, we also measure the variation of each app's rank within its peer groups and globally. First, for each studied app, we calculate its monthly rating, i.e., the average rating of the reviews of an app on a monthly basis. Second, for each month, we calculate the rank of each app within its peer group based on their average ratings in that month (i.e., the within-peer-group rank of an app). For each month, we also calculate the rank of each app within a random group that is created by randomly drawing 10 apps from the 100 studied apps (i.e., the global rank of an app). We use a percentile rank (ranges from 0% to 100%), where 0% means the highest rank while 100% means the lowest rank. Finally, we calculate the standard deviation of each app's monthly rank within its peer group and random group, separately. The standard deviation captures how an app's monthly rating relative to other apps change over time.

We use the Wilcoxon signed-rank test to evaluate the statistical difference between the standard deviation of the studied apps' monthly ranks within peer groups and globally (i.e., within random groups). We use the Wilcoxon signed-rank tests instead of the paired t-test as the standard deviation of the monthly ranks of the studied apps is not normally distributed (i.e., the Shapiro-Wilk test shows a p-value $< 0.05$).

*3) Results:* **Comparing app ratings within peer groups provides a new perspective for comparing app ratings. For instance, 4.5 is a low rating for Bible and Security apps while it is a high rating for many other peer groups.** Our Kruskal-Wallis test shows that the average ratings of the studied apps are statistically different between peer groups (i.e., p-value $< 0.05$). Figure 2 uses box plots to show the distribution of the average app ratings of the studied apps which are captured at the end of the studied period. As we select popular apps in the Google Play Store, most of the apps have an average rating higher than four. However, we still observe that users tend to assign high ratings for apps from some peer groups (e.g., the Bible group) while assigning lower ratings for apps from other peer groups (e.g., the Free call group). For example, even the studied Bible app with the lowest average rating has a higher rating than all the studied Free call apps. Some peer groups (e.g., the Browser, News and SMS groups) have a wider distribution of app ratings, while the app ratings in some other peer groups (e.g., the Weather, Security and Bible groups) are very consistent.

We believe that users have different standards for apps from different peer groups, because (1) apps from different peer groups offer different functionalities, and (2) users may require (or expect) higher quality for certain functionalities. For example, the fact that the Free call apps have relatively low ratings might be explained by the assumption that users expect a higher quality of the Free call apps, since the failures of these apps might interrupt users' important communications.

**The peer analysis of app ranks provides a more relevant view about how an app rank changes over time than the global analysis, as we observe that the rank of an app within its peer group is more subject to change than its global rank.** Our Wilcoxon signed-rank test shows that the standard deviation of an app's monthly ranks within its peer group is statistically different from the standard deviation of its monthly ranks globally (i.e., p-value $< 0.05$). Figure 3 compares the standard deviation of an app's percentile rank in its peer group and globally over time (per month). As shown

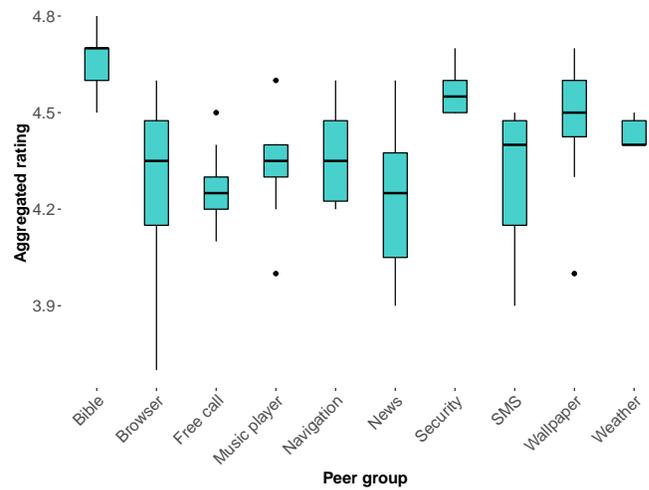

Fig. 2. Distribution of the average app ratings at the end of the studied period. Each boxplot shows the distribution of the average ratings of the ten apps in one peer group.

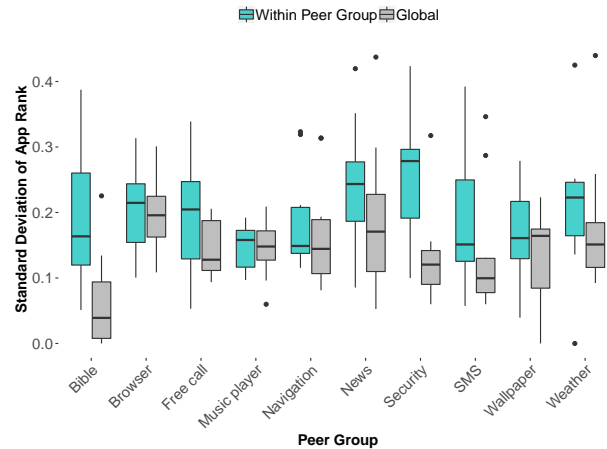

Fig. 3. Comparing the standard deviations of peer and global ranks of apps over time (per month). Each boxplot shows the distribution of the standard deviation of each peer app' ranks over time. Each line within a box plot shows the median standard deviation.

in Figure 3, the standard deviation of an app's rank in its peer group is generally higher than the standard deviation of its ranks globally. For example, the median standard deviation of the ranks of the security apps within their peer group is around 30%, which indicates a ∼30% rank change over time. In contrast, the median standard deviation of the global ranks is only around 10%.

> **Summary of RQ 1**
>
> We observe that comparing user ratings within peer apps can provide very different results from comparing user ratings from a global perspective. App store owners may consider providing an additional rating mechanism that normalizes app ratings within peer groups, to provide app developers and users a different perspective about the position of an app.

*B. RQ2: How does the peer analysis of user reviews differ from the global analysis?*

*1) Motivation:* Prior research analyzes user reviews of apps to extract useful information about users' feedback, such as users' perception about app features or user-reported bugs [3, 37, 38]. These studies usually analyze user reviews in general instead of focusing on apps from the same peer group. In this RQ, we want to understand whether an analysis of user reviews within peer groups can provide insights about the users' main concerns of the peer apps that are different from the general analysis of user reviews.

*2) Approach:* **General topics and app-specific topics.** After our topic modeling and manual card sorting, we generate 46 high-level topics. There are two topics which are about "good apps" and "bad apps", through which users express how they like the apps without any actionable reason. Therefore, we remove these two topics from our analysis. Among the remaining 44 topics, we found 15 topics that are specific to apps in specific peer groups, such as the "navigation" topic to the Navigation apps. The other 29 topics are general across peer groups, such as the "bug" and "UI design" topics. We name them *app-specific topics* and *general topics*, respectively. **Topic assignment.** In order to quantitatively understand users' concerns about their apps, we use the topic assignment [39] to measure the total presence of a topic in a set of user reviews. As the number of reviews varies from app to app, we further define an **average topic assignment (TA)**[2] metric to measure the importance of each topic in the studied reviews. The TA of a topic $z_i$ measures the average presence of the topic in a number of reviews, and it is defined as

$$TA(z_i) = (\sum_{j}^{N} \theta_{ij})/N \qquad (1)$$

where N is the number of considered reviews (e.g., the reviews of a peer group) and $\theta_{ij}$ the membership of topic $z_i$ in the $j$th review. A larger TA of a topic means that a larger portion of the considered reviews is related to the topic. The TA values of all the topics sum up to 1 for the considered reviews. When we calculate the TA of a topic group that represents multiple original topics, we use a sum of their $\theta$ values which represents the combined membership of the original topics in a review. **Standard Deviation (SD) of TA.** We also measure the standard deviation (SD) of each topic in each peer group and across peer groups. Our intuition is that the topic assignment may have smaller SD within peer groups (i.e., homogeneity) and bigger SD across peer groups (i.e., heterogeneity). For each topic, we calculate its topic assignment for each studied app. Then, we calculate the SD of its assignment in the apps within each peer group. For each topic, we also calculate the SD of its assignment in each random group (we create 10 random groups by randomly assigning 10 apps to each group). Finally, we use the Wilcoxon rank-sum test to evaluate the statistical difference between the SD of topic assignment within peer groups and the SD of topic assignment within random groups (i.e., across peer groups, or globally).

[2]In this paper, we use the terms "topic assignment" or "TA" to represent the average topic assignment.

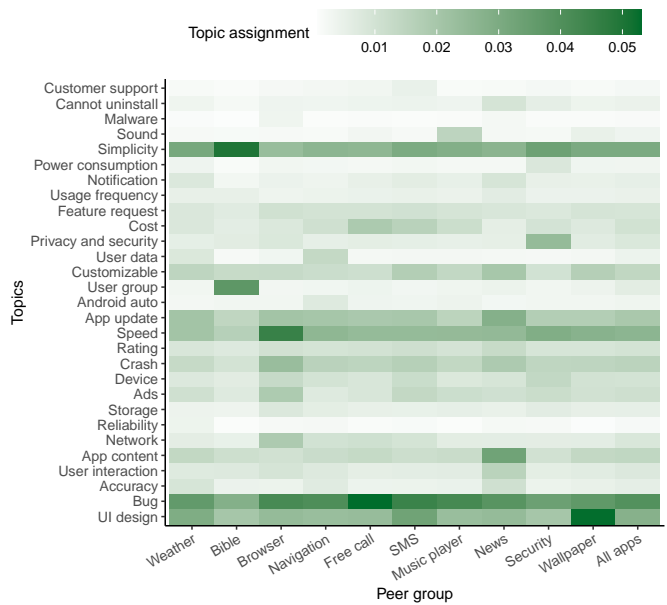

Fig. 4. A heat map that shows the topic assignment of general topics in different peer groups and in all studied apps.

*3) Results:* **Peer-group analysis provides a complementary perspective to spot the dominant topics in the user reviews of a peer group.** Table IV shows the top ten mentioned topics in each peer group and in all the apps combined together (i.e., globally). We use the TA of each topic in the reviews of each peer group to rank the top mentioned topics groups. When considering all the apps together (i.e., as shown in the "All apps" column of Table IV), the general topics of "bug", "simplicity", "UI design", and "speed" are the most frequently mentioned topics. However, for seven out of the ten studied peer groups, the app-specific topics (marked as bold), such as the "weather features" topic for the Weather apps, are the most frequently mentioned topics. Some topics are among the most frequently mentioned topics in certain peer groups (e.g. the "user group" topic for the Bible group, and the "crash" topic for the Browser group); while the same topics are missing from the top ten topics from a global view.

**General topics (e.g., "speed" and "UI design") are mentioned heterogeneously in the reviews of the apps across peer groups.** Figure 4 visualizes the topic assignment of the general topics across peer groups. The topic "speed" is the most frequently mentioned topic in the Browser group; however, the same topic is much less important globally (i.e., as shown in the "All apps" column) and in other peer groups such as the News group. Browser app users are more concerned about speed than the users of other apps. The topic "bug" is the most frequently mentioned topic in general and in most peer groups, especially for the Free call apps. As we discussed in the previous RQ, users of the Free call apps might care more about the failures of such apps, since failures of such apps might interrupt instant communications that are

TABLE IV
TOP TEN MENTIONED TOPICS IN EVERY PEER GROUP AND IN ALL STUDIED APPS (RANKED BY THE TOPIC ASSIGNMENT). THE TOPICS THAT ARE HIGHLIGHTED IN BOLD ARE APP-SPECIFIC TOPICS.

| Weather | Bible | Browser | Navigation | Free call | SMS | Music player | News | Security | Wallpaper | All apps |
|---|---|---|---|---|---|---|---|---|---|---|
| **Weather features** | **Bible features** | Speed | **Map features** | Bug | **Messaging features** | **Multimedia** | **News content** | **Antivirus** | UI design | Bug |
| Bug | Simplicity | Bug | Bug | **Calling features** | Bug | Bug | **News source** | Bug | Bug | Simplicity |
| Simplicity | User group | **Web browser** | Simplicity | **Messaging features** | UI design | Simplicity | Bug | Simplicity | Simplicity | UI design |
| UI design | Bug | **Multimedia** | Speed | Simplicity | Simplicity | Speed | App content | Speed | Speed | Speed |
| Speed | **Multimedia** | UI design | UI design | Speed | Speed | UI design | App update | **Cleaner** | **Wallpaper** | **Weather features** |
| App update | UI design | Simplicity | App update | UI design | App update | App update | Simplicity | Privacy and security | App update | **Multimedia** |
| Customizable | Speed | Crash | Crash | **Free call apps** | Customizable | Sound | Speed | UI design | Customizable | **Messaging features** |
| App content | App update | App update | **Weather features** | App update | Crash | Crash | UI design | App update | Crash | **Bible features** |
| Crash | Customizable | Network | User data | Cost | Cost | Customizable | Customizable | Crash | App content | App update |
| **News content** | App content | Ads | App content | Crash | **Calling features** | App content | Crash | Device | **Multimedia** | **Map features** |

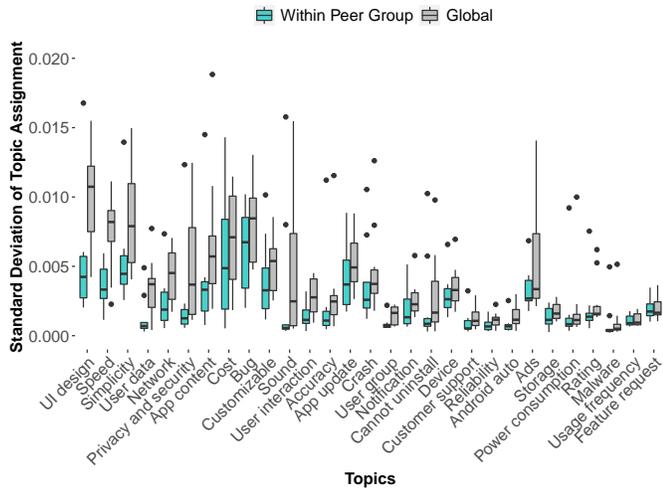

Fig. 5. The distributions of the standard deviation of the `general topics`' assignment within peer groups vs. within random groups (i.e., globally).

very important to users. However, the topic "bug" is less of a concern in Bible apps. Instead, the topics of "user group" and "simplicity" are more concerned in Bible apps. The topic "user group" is rarely mentioned in general or in any other peer group. Bible users care more about user groups, such as kids and adults. For example, users of the "Superbook Bible, Video & Games" app wrote a review *"Great for kids and adults to read bible"*.

While the topic of "bug" is generally more frequently concerned than the topic of "UI design", "UI design" is more frequently mentioned than the "bug" topic in the Wallpaper apps. Some topics are only concerned in a few app groups, such the topics of "app content" and "cannot uninstall" to News apps, "sound" to MusicPlayer apps, "user group" to Bible apps, and "network" to Browser apps. The `heterogeneity` of review topics in different peer groups suggests future studies on user reviews to pay more attention on analyzing review topics within peer groups.

**General topics are mentioned more homogeneously in the reviews of the apps within peer groups.** Our statistical test shows that the SD of the topic assignment within peer groups is smaller than the SD of the topic assignment within random groups in a statistically significant manner (i.e., our Wilcoxon rank-sum test shows a p-value smaller than 0.05), which means the topics present more homogeneously within peer groups than across peer groups. Figure 5 compares the distributions of the SD of the general topics' TA within peer groups and random groups (i.e., globally). As shown in Figure 5, 97% (28 out of 29) of the topics' TAs are more consistent within peer groups (i.e., with a smaller median SD) than within random groups, which concurs our Wilcoxon rank-sum test results (i.e, p-value < 0.05). For example, the median SD of the topic "UI Design" is 0.005 within peer groups while it is 0.011 within random groups (i.e, more than two times larger). Such results indicate that users tend to have more consistent concerns for apps within peer groups.

> **Summary of RQ 2**
>
> Peer-group analysis provides a different perspective to spot the dominant topics in the user reviews of a peer group. For example, a general topic that is critical for one peer group can be much less important in other peer groups or from a global perspective.

*C. RQ3: How do review topics contribute to the negative ratings within peer apps versus globally?*

*1) Motivation:* Users usually post reviews and assign ratings to their downloaded apps. The ratings often indicate users' satisfaction about different aspects (i.e., review topics) of an app as expressed in the corresponding reviews. In previous RQs, we find the heterogeneity of users' reviews and ratings for apps from different peer groups. In this RQ, we want to understand how review topics contribute to the ratings of an app, from both a within-peer-group and a global perspectives.

In particular, we want to understand whether a within-peer-group analysis can provide different perspectives about how review topics contribute to app ratings in each peer group.

*2) Approach:* In this RQ, we associate a topic with the rating of each review that contains that topic to study how the topic contributes to user ratings. As a review about a topic (e.g., "speed") can be either a positive review (e.g. *"Awesome browsing speed'*) or a negative review (e.g., *"Slow speed"*), it is misleading to study the average contribution of a topic on user ratings. Prior work suggests that negatives reviews are usually more informative than positive reviews, as negative reviews usually directly indicate that users do not like certain characteristics of an app [3, 40]. Therefore, in this RQ, we analyze the negative contribution of the review topics. We follow prior work [3, 4, 26] and classify ratings with less than three stars as low (i.e., negative) ratings, as prior work shows that users usually will not download an app with less than three stars [41].

**Negative contribution of topics.** We define a **negative contribution** (**NC**) metric to measure the negative contribution of a topic on app ratings. The NC of a topic $z_i$ is calculated as

$$NC(z_i) = (\sum_{\substack{j \\ r_j <= 2}}^{N} \theta_{ij}) / (\sum_{\substack{j \\ r_j <= 2}}^{N} 1) \quad (2)$$

where N is the number of considered reviews, $r_j$ is the star rating of the $j$th review, and $\theta_{ij}$ is the membership of topic $z_i$ in the $j$th review. In this equation, we only consider the reviews with one or two stars as our goal is to evaluate a topic's contribution to the negative reviews. The NC of a topic is actually the proportion of negative reviews (i.e., reviews with one or two stars) that are contributed by the topic. NC ranges from 0 to 1, a larger value indicates a bigger negative contribution. For example, a NC value of 0.1 means that the topic contributes to 10% of the negative reviews.

**Standard Deviation (SD) of NC.** Similar to RQ2, we measure the SD of each topic's NC within peer groups and across peer groups. We use the Wilcoxon rank-sum test to evaluate the statistical difference between the SD of each topic's NC within peer groups and the SD of each topic's NC across peer groups (i.e., within random groups). Our assumption is that the NC of the topics may have smaller SD within peer groups (i.e., homogeneity) and bigger SD across peer groups (i.e., heterogeneity).

*3) Results:* **A global review analysis can hide the app aspects that contribute the most negative reviews in some peer groups.** Table V lists the top ten negative topics in each peer group and in all the studied apps (i.e., globally). We use the NC metric to rank the topics. The general topics of "bug", "app update", and "crash" have the largest negative contribution in all the apps combined together. However, app-specific topics have the largest negative contribution in five of the ten peer groups. The topic of "Ads" is among the most negative-contributing topics in the Bible and Security groups; however, the same topic contributes to much less negative reviews in other peer groups and globally. Similarly, the topic of "user data" is one of the topics that contribute to the most negative reviews in the Navigation group; however, the same topic is hidden from the top ten negative topics of other peer groups (except for the Weather group) and globally.

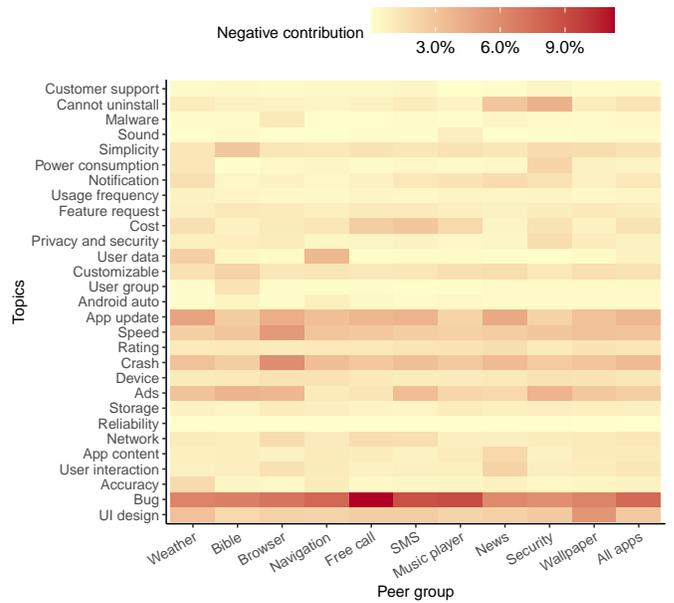

Fig. 6. A heat map that shows the negative contribution of `general topics` in different peer groups and in all studied apps.

**A general topic that appears less "harmful" in some peer groups or from a global perspective can actually bring much more negative contribution to certain peer groups.** Figure 6 shows the negative contribution of each general topic across all apps and in each peer group. Even though the topic "bug" is the most negative general topic for each peer group, it has a much higher negative contribution in the Free call peer group than in any other peer group. The topic of "UI" has a more negative contribution in the Weather and Wallpaper peer groups than in other peer groups. The topics of "crash" and "speed" have a higher negative contribution in the Browser apps than in other peer groups. The topic of "Ads" is more negatively contributing in some peer groups (e.g., Browser apps) and much less contributing in other peer groups (e.g., Navigation apps). Some topics are only negatively contributing in certain peer groups, such as the topic of "cannot uninstall" to the News and Security peer apps. The difference of the topics' negative contribution across peer groups (i.e., `heterogeneity`) suggest future work to study how different aspects contribute to the ratings/ranks of apps within peer groups.

**While general topics have heterogeneous negative contributions across peer groups, these topics present more homogeneous negative contributions within peer groups.** We find that the SD of each topic's NC within peer groups is smaller than the SD of each topic's NC across peer groups (i.e., within random groups) in a statistically significant manner (i.e., our Wilcoxon rank-sum test shows a p-value less than 0.05), which means the NC of the topics are more consistent

TABLE V
TOP TEN NEGATIVE TOPICS IN EVERY PEER GROUP AND IN ALL THE STUDIED APPS (RANKED BY THE NEGATIVE CONTRIBUTION). THE TOPICS THAT ARE HIGHLIGHTED IN BOLD ARE THE APP-SPECIFIC TOPICS.

| Weather | Bible | Browser | Navigation | Free call | SMS | Music player | News | Security | Wallpaper | All apps |
|---|---|---|---|---|---|---|---|---|---|---|
| **Weather features** | **Bible features** | Bug | **Map features** | Bug | **Messaging features** | **Multimedia** | **News content** | Bug | Bug | Bug |
| Bug | Bug | Crash | Bug | **Calling features** | Bug | Bug | Bug | Cannot uninstall | UI design | App update |
| App update | Ads | Speed | User data | App update | App update | Crash | **News source** | Ads | App update | Crash |
| UI design | Speed | App update | Crash | **Messaging features** | Ads | Speed | App update | **Antivirus** | Speed | **News content** |
| Crash | Simplicity | Ads | App update | Crash | Crash | App update | Crash | Speed | Crash | Speed |
| Ads | **Multimedia** | **Multimedia** | Speed | Speed | Cost | UI design | Cannot uninstall | UI design | Ads | UI design |
| Speed | App update | **Web browser** | UI design | Cost | UI design | Ads | Speed | Crash | **Wallpaper** | **Messaging features** |
| User data | Crash | UI design | Device | UI design | Speed | Cost | UI design | Cleaner | Simplicity | Ads |
| Accuracy | Customizable | Network | Customizable | **Free call apps** | **Calling features** | Customizable | User interaction | App update | Customizable | **Multimedia** |
| Notification | UI design | **Messaging features** | Cost | Network | Network | Simplicity | App content | Power consumption | Device | **Map features** |

## IV. THREATS TO VALIDITY

**External Validity.** This work selects 100 apps across ten peer groups as our subject apps. Some of our results (e.g., the review topics) may not generalize to apps in other peer groups. In order to reduce such limitation, this work selects ten peer groups across a broad range of app categories. In addition, this work may not cover all the peer apps in the studied peer groups. Instead, we choose ten apps for each peer group so that our analysis is not biased by peer groups that have larger numbers of apps. The purpose of our paper is to demonstrate that studying apps from the perspective of peer apps (where such a peer group is created in a sensible manner) can provide useful insights about the characteristics of peer groups. We can always add more peer groups and peer apps, but the message and findings will be the same – peer-app analysis can provide a unique and important perspective to understanding the characteristics (e.g., user ratings and critical topics) of apps.

**Internal Validity.** In RQ3, we analyze the negative contribution of review topics to app ratings by relating the review topics with the ratings in the same user reviews. In particular, we analyze the review topics that are associated with low ratings. However, the review topics might not indicate the reason that a user provides a low rating to an app. Besides, different users may have different standard for "high" or "low" ratings. In this work, following prior work [3, 4, 26], we classify ratings with one or two stars as low (i.e., negative) ratings, as prior work shows that users usually will not download an app with less than three stars [41]. Future study can re-explore our observations through user studies to understand users' rationale behind assigning low ratings to apps.

**Construct Validity.** In order to demonstrate the importance of performing peer-app analysis, we hand-selected peer apps to form peer groups. We read the app title and description of the top 2,000 popular apps and identified peer apps that provide similar major functionalities and grouped them into peer groups. Our selection results may be biased by the individuals within peer groups than across peer groups. Figure 7 compares the distributions of the SD of the general topics' NC within peer groups and within random groups. Figure 7 shows that 79% (23 out of 29) of these topics' NC is more consistent within peer groups (i.e., with a smaller median SD) than within random groups. For example, the median SD of the topic "User data" is 0.003 within peer groups and 0.012 within random groups (i.e., four times larger). However, there are a few exceptions (e.g., the "Cost" topic) that show as inconsistent NC within peer groups as within random groups.

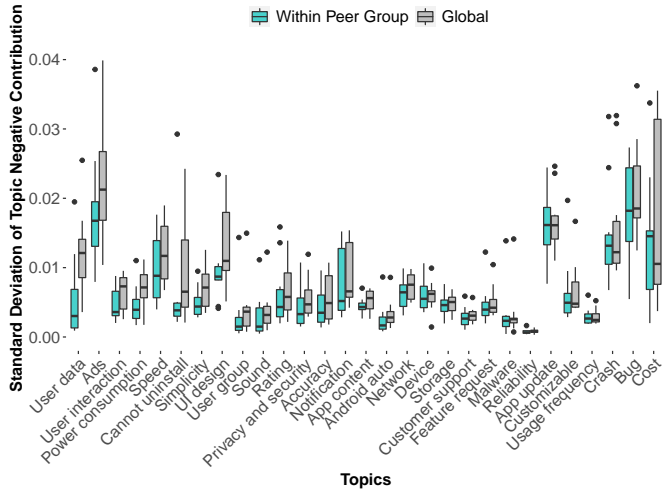

Fig. 7. The distributions of the standard deviation of the `general topics`' negative contribution within peer groups vs. random groups (i.e., globally).

> **Summary of RQ 3**
>
> Analyzing the contribution factors of app ratings from the perspective of peer apps can produce more relevant observations for each peer group. A general topic that appears less "harmful" from a global perspective can bring more severe negative impact to certain peer groups.

who performed the manual selection process. Nevertheless, we expect that the actual developers of an app are the only ones who are truly capable of determining their peer apps (i.e., competitors). Besides, identifying peer apps and peer group is not the main goal of this work.

In this work, we use a combination of automatic topic modeling and manual coding to extract topics from app reviews. Determining the appropriate number of topics is usually a subjective process. Besides, existing approaches for determining the optimal number of topics are usually heuristic-based; as shown in prior work [26], different approaches (e.g., [27] and [28]) can produce very different optimal numbers of topics. In this work, instead, we spend significant manual effort to analyze the automatically generated topics. We first run LDA using a relatively large number of topics (i.e., 500 topics), as suggested by prior work [24, 29, 30]. Then, we perform a card sorting process to manually group similar topics together. Such a combination helps us identify more meaningful topics than only running LDA with a smaller number of topics [30].

In this work, we apply topic modeling on a corpus of reviews of all the studied apps combined together. Applying topic modeling within each peer group may provide better topics that are relevant to the apps within that peer group. In this work, however, we need to compare the extracted topics across peer groups. Building separate topic models for each peer group can make it hard to compare the topics generated for different peer groups (i.e., different topic models have different set of topics). Therefore, we instead extract the topics of the reviews of the studied apps using a single topic model. The generated topics, which are shared by different peer groups, allow us to compare the distribution of these topics across peer groups.

## V. RELATED WORK

In this section, we discuss prior work related to analysis of user reviews and analysis of peer apps.

**Analyzing User Reviews.** Prior work proposes approaches that automatically classify (e.g., using Naive Bayes) user reviews into a few number of categories (e.g., feature requests, or bug reports) [2, 37, 42, 43]. Prior work also extracts topics from user reviews to help app developers better understand user reviews [1, 44–47]. For example, Chen et al. [45] propose AR-Miner (Automatic Review Miner) that filters out non-informative reviews and groups similar reviews together based on topic extraction. AR-Miner ranks topics based on different criteria such as the number of reviews containing a topic and the average rating of a topic. AR-Miner is useful for app developers to identify user-raised topics over time and identify reviews that are related to a certain topic.

Prior research mainly focuses on analyzing reviews of apps that are distributed across different app categories. In this work, we demonstrate the benefit of the peer-group-level analysis of user reviews to better understand the apps ratings and the critical topics in each peer group. Hence, we encourage software engineering researchers and tool developers to pay more attention to peer-app analysis, as such analysis will help app developers better understand the characteristics of specific peer groups and prioritize their efforts.

**Analyzing Peer Apps.** Prior work shows that app developers care about comparing the characteristics of their apps (e.g., ratings) with their competitor apps that provide similar functionalities [48]. Since app categories contain a broad range of apps [8], developers need to compare their apps against a small group of closely related apps. Hence, app stores such as the Google Play Store recently enabled developers to compare their app with a custom-defined peer group that contains closely related apps [6].

Prior work proposed different approaches to identify closely related apps (i.e., peer apps) in app stores [7–10, 49] (e.g., based on app descriptions [8] or user reviews [9]). Prior work also proposed approaches that aim to help app developers improve their apps using the characteristics of closely related apps [9, 10, 26, 50–53]. For example, Nayebi et al. [50] and Jiang et al. [10] extract features from the descriptions of peer apps. Then, they prioritize the features that need to be included in the next releases based on the importance (e.g., the frequency and the ratings) of such features in peer apps. Noei et al. [26] study 4,193,549 user reviews of 623 apps in the Google Play store. Noei et al. identify the key topics (i.e., the most frequently mentioned topics) in every app category. Noei et al. find that the release notes of the highly-rated releases have a significant correlation with the key review topics of the app categories.

Our work confirms and extends prior work by demonstrating the importance of performing peer-app analysis, with scientific evidence. Through an experiment of analyzing app ratings, review topics, and the impact of review topics, our work shows that peer-group analysis provides a unique and important perspective to understanding the app ratings and the dominant or influential aspects of apps in a peer group.

## VI. CONCLUSIONS

Peer apps provide similar functionalities (e.g., weather forecasting) and they are direct competitors to each other. In this work, we show the importance of performing peer-app analysis by studying 100 apps across ten peer groups. Through analyzing the ratings and review topics of these 100 apps over a period of 21 months, we show that performing peer-app analysis can provide a unique and more relevant view than performing app analysis from a global perspective. For example, a general review topic that is critical for one peer group can appear much less important in other peer groups or from a global perspective, and a seemly "harmless" topic from a global perspective can be much more "harmful" to certain peer groups. Our findings motivate future efforts to contextualize their work from the perspective of peer apps, to provide more relevant support for the development of apps in specific peer groups.